\documentclass[prl,aps,twocolumn,preprintnumbers,showpacs,superscriptaddress]{revtex4}
\usepackage[colorlinks=true, pdfstartview=FitV, linkcolor=red, citecolor=blue, urlcolor=blue]{hyperref}

\usepackage{graphics,graphicx,epsfig,bm}

\usepackage{epstopdf}

\begin{document}

\title{Thermal Noise on Adiabatic Quantum Computation}

\date{\today}

\author{Man-Hong Yung}
\email[email: ]{myung2@uiuc.edu}

\affiliation{Department of Physics, University of Illinois at
Urbana-Champaign, Urbana IL 61801-3080, USA}


\begin{abstract}
The success of adiabatic quantum computation (AQC) depends crucially on the ability to maintain the quantum computer in the ground state of the evolution Hamiltonian. The computation process has to be sufficiently \textit{slow} as restricted by the minimal energy gap. However, at finite temperatures, it might need to be \textit{fast} enough to avoid thermal excitations. The question is, how fast does it need to be? The structure of evolution Hamiltonians for AQC is generally too complicated for answering this question. Here we model an adiabatic quantum computer as a (parametrically driven) harmonic oscillator. The advantages of this model are (1) it offers high flexibility for quantitative analysis on the thermal effect, (2) the results qualitatively agree with previous numerical calculation, and (3) it could be experimentally verified with quantum electronic circuits.
\end{abstract}

\maketitle
Adiabatic quantum computation (AQC) offers an alternative route for achieving computational goals \cite{Farhi00}, compared with the ``standard model" of quantum computation based on the gate model. The basic idea of AQC is very simple: to maintain the system (computer) to stay at the (assumed unique) ground state with respect to a time-dependent Hamiltonian. For an isolated quantum system, this is in principle guaranteed by the quantum adiabatic theorem for sufficiently \textit{slow} evolution, depending on the energy gap between the (instantaneous) ground state and the first excited state. Practically, AQC would be operated under some finite temperature which may not necessarily be negligible (recall energy gaps usually shrink with the increase of the problem size) compared with the energy gap. The effect of thermal noise would then play an important role in determining the performance of AQC. Physically, the relaxation process (i.e. excitation to higher energy states) takes finite time to complete. Hence, we expect that AQC needs to be sufficiently \textit{fast} as well. This sets another time scale due to the environment. Consequently, unless thermalization is not an issue, \textit{AQC would work only if the computation time lies within these two time scales}.  The question is, how to determine the latter time scale involved? We shall answer this question in this letter.

This work is motivated by recent studies \cite{Childs01,Shenvi03, Roland05, Sarandy05,Aberg05, Ashhab06, Tiersch07,Amin08} related to the question of robustness of AQC. Concerning the noise effect on AQC, some of the models are based on either qualitative or perturbative arguments which are not verified by independent numerical investigation. Some of them are formulated in terms of parameters which are inaccessible experimentally. On the other hand, it was believed \cite{Ashhab06,Tiersch07,Amin08} that two-level approximation would be valid for AQC, even if large number of excited states would be involved when the minimal gap is smaller than the temperature. It is therefore still unclear how ``robust" AQC is against thermal noise.  

With these problems in mind, our goal here is to study the thermalization problem of AQC by modeling a Harmonic oscillator as a quantum computer. This model not only provides us with enough (infinite) excited states but also allows \textit{quantitative} analysis. As we shall see, it could not be modeled by the two-level approximation. Moreover, we shall quantify the effects on the performance of AQC through physical quantities such as temperature $T$, relaxation time $1/ \gamma$, energy gap $\Delta$ and computation time $\tau_*$. The ``anomaly" of this model may seem to be the evenly distributed energy levels. To verify the validity, we have compared it with the numerical simulation by Childs \textit{et al.} \cite{Childs01}, and found that the predictions of this harmonic model qualitatively agrees with their results. Lastly, this model is testable with the current quantum electronic technologies, e.g. simple RLC circuits.

\emph{Adiabatic Quantum Computation ---} To define our adiabatic quantum computer, there are only two adjustable and time-varying parameters, namely the ``mass" $m_t  \equiv m\left( t \right)$ and the ``spring constant" $k_t  \equiv k\left( t \right)$. The time dependence of these two parameters, at this stage, is completely arbitrary and is designed to \textit{simulate} (e.g. see example III below) an adiabatic quantum computer. The (computational) system Hamiltonian $H_S(t)$ is described by that of a standard parametrically driven harmonic oscillator:
\begin{equation}\label{HS}
H_S \left( t \right) = \frac{{{\hat p}^2 }}{{2m_t }} + \frac{1}{2}k_t {\hat x}^2 \quad,
\end{equation}
which is associated with a set of (instantaneous) energy eigenstates $ \left| {n_t } \right\rangle$ , with $n = 0,1,2,...$, satisfying the eigenvalue equation: $H_S \left( t \right)\left| {n_t } \right\rangle  = E_n \left( t \right)\left| {n_t } \right\rangle$. Here $E_n \left( t \right) = \left( {n + 1/2} \right)\Delta \left( t \right)$ is the  (instantaneous) eigen-energy for the state $\left| {n_t } \right\rangle$. The energy gap $\Delta \left( t \right) \equiv \sqrt {k_t /m_t }  = E_{n + 1} \left( t \right) - E_n \left( t \right)$ does not depend on $n$, by definition. The initial state is assumed to be the ground state $\left| {0_{t = 0} } \right\rangle $ of $H_S \left( t = 0 \right)$. In the absence of the heat bath, the final state is given by $U\left( {t = t_f } \right)\left| {0_{t = 0} } \right\rangle$, where $U\left( t \right) = T\exp ( { - i\int_0^t {H_S\left( {t'} \right)dt'} })$ (with $\hbar  = 1$) is a time-ordered series. 

The computation is considered to be fail if the final state deviates significantly (due to excitation to higher energy states) from the desired ground state $\left| {0_{t = f} } \right\rangle$ of $H_S \left( {t = t_f } \right)$. This is best quantified by the fidelity $ F \equiv \left| {\left\langle {0_{t = f} } \right|U\left( {t_f } \right)\left| {0_{t = 0} } \right\rangle } \right|^2$. Here since our goal is to study the thermal effect from the environment, we assume that AQC in the absence of the heat bath can be achieved (almost) perfectly, i.e., $F \approx 1$; violation of this condition may be considered as perturbation.

Under this condition (and to zeroth order in $\dot \Delta \left( t \right)$), we may write $U\left( t \right)\left| {n_0 } \right\rangle  = \exp ( { - i\int_0^t {E _n \left( {t'} \right)dt'} })\left| {n_t } \right\rangle$, and hence a relation which is needed later:
\begin{equation}\label{Inter_pict}
U^\dagger ( t )a_t U\left( t \right) = \exp \left( { - i\int_0^t {\Delta \left( {t'} \right)dt'} } \right)a_0 \quad,
\end{equation}
where $a_t  \equiv \sqrt {m_t \Delta _t /2} \left( {{\hat x} + i{\hat p}/m_t \Delta _t } \right)$ is the (instantaneous) annihilation operator for $H_S(t)$. 

\emph{Ground State Occupation ---} In the presence of a heat bath, a mixed state representation $\rho \left( t \right)$, or the reduced density matrix $\rho _S \left( t \right) = Tr_B \left\{ {\rho \left( t \right)} \right\}$, is needed. The performance of the quantum computer is determined by the ground state occupation $P_g  \equiv \left\langle {0_t } \right|\rho _S \left( t \right)\left| {0_t } \right\rangle$, and in the coordinate space
\begin{equation}\label{Pg_gen}
P_g  = \int {\int_{ - \infty }^\infty  {dx'dx} } \left\langle x \right|\rho _S \left( t \right)\left| {x'} \right\rangle \varphi _t^* \left( x \right)\varphi _t \left( {x'} \right) \quad,
\end{equation}
where $\varphi _t \left( x \right) \equiv \left\langle x \right|\left. {0_t } \right\rangle  = \left( {m_t \omega _t } \right)^{1/4} \exp \left( { - m_t \omega _t x^2 /2} \right)$ is the (instantaneous) ground state wavefunction of $H_S \left( t \right)$. Before going into the technical details of the calculations for $\rho _S \left( t \right)$, we first argue that, subject to the constraints (a), (b) and (c) described below, the relevant quantity here is only the physical observable $\langle {\hat x\left( t \right)^2 }\rangle  = Tr\left\{ {{\hat x}^2 \rho \left( t \right)} \right\}$ (or the current fluctuation $\left\langle {I(t)^2 } \right\rangle$ in RLC circuits).

The imposed constraints are (a) the heat bath can be approximated by a set of harmonic oscillators $H_B  = \sum\nolimits_k {\hbar \omega _k } b_k^\dagger b_k$, (b) the system-bath coupling $H_{SB}$ is bilinear e.g. terms like $\hat x ( {b_k  + b_k^\dagger  })$, and (c) the initial state of the bath is in a thermal state $\rho _B  = e^{ - \beta H_B } /Tr\left[ {e^{ - \beta H_B } } \right]$ and the system is in the ground state of $H_S \left( 0 \right)$, i.e., $\rho \left( 0 \right) = \left| {0_{t=0} } \right\rangle \left\langle {0_{t=0} } \right| \otimes \rho _B$. To proceed, we write
\begin{equation}\label{outer}
\left| {x'} \right\rangle \left\langle x \right| =  \frac{1}{{2\pi }}\int_{ - \infty }^\infty  {d\nu } e^{i\nu \left( {\mu /2 - x} \right)} e^{i\left( {\mu \hat p + \nu \hat x} \right)} \quad ,
\end{equation}
where $\mu  \equiv x - x'$ (and $\nu $ is a just dummy variable). This form suggests that we have to evaluate the quantity $\left\langle {e^{i\left( {\mu \hat p + \nu \hat x} \right)} } \right\rangle  = Tr\left\{ {e^{i\left( {\mu \hat p + \nu \hat x} \right)} \rho \left( t \right)} \right\}$, which is equal to $\exp ( { - \langle {( {\mu \hat p + \nu \hat x} )^2 } \rangle /2} )$ from the Bloch identity. Now, as verifiable by the master equation, here we claim that $\left\langle {a_t^2 } \right\rangle  = \langle {a_t^{ \dagger 2} }\rangle  = 0$. By completing the Gaussian integrals in Eq. (\ref{Pg_gen}) and (\ref{outer}), we finally arrive at a very compact form for $P_g$:
\begin{equation}\label{compact_Pg}
P_g  = \frac{1}{{1 + n\left( t \right)}} \quad,
\end{equation}
where $n\left( t \right) \equiv \langle {a_t^\dagger  a_t } \rangle  = Tr \{ {a_t^\dagger  a_t \rho \left( t \right)} \}$. Thus, as advertised, $\left\langle {x^2 } \right\rangle  = \left( {\hbar /2m_t \Delta_t } \right)( {2\langle {a_t^ \dagger  a_t } \rangle  + 1} )$ is the only quantity needed to determine $P_g$.

\emph{Master Equation ---} We shall obtain $n(t)$ through the master equation approach \cite{Carmichael}. Here the full Hamiltonian is divided into three parts: $H = H_S \left( t \right) + H_B  + H_{SB}$ where the first two terms have been defined. We assume that the coupling term $H_{SB}$ is a time independent operator (i.e. indepdendent of the mass $m_t$ and spring constant $k_t$ of the oscillator), and is explicitly given by 
\begin{equation}
H_{SB}  = {\hat x}\sum\limits_k {g_k \left( {b_k^ \dagger   + b_k } \right)} \quad.
\end{equation}
Note that there could be a frequency renormalization (lamb shift type), which modifies the ground state wavefunction. This effect would be small for weak damping $\Delta \left( t \right) \gg \gamma \left( t \right)$ where $\gamma \left( t \right) \equiv \eta \left( t \right)/m\left( t \right)$ (cf. Eq.(\ref{damped}) and (\ref{nt})). Second, even for ohmic damping (time independent $\eta \left( t \right) = \eta$), the relaxation rate $\gamma \left( t \right) \propto 1/m\left( t \right)$ (or time to reach equilibrium) depends on the system parameter (here the ``inertia" $m(t)$), and therefore may be time-dependent.

To continue, we shall keep the standard assumptions for the master equation, namely (i) product initial state, (ii) Born-Markov approximation (i.e. weak coupling and short memory time), and (iii) rotating wave approximation (i.e. ignore fast oscillations). Subject to these constraints, the master equation is given \cite{Carmichael} by
$$
\frac{d}{{dt}}\tilde \rho _S  = \frac{{ - 1}}{{\hbar ^2 }}\int_0^t {dt'} Tr_B \{ {[ {\tilde H_{SB}( t),[ {\tilde H_{SB}( {t'} ),\tilde \rho _S( t) \otimes \rho _B } ]} ]}\} ,
$$
where in the interaction picture: $\tilde \rho _S( t ) \equiv U^ \dagger ( t )\rho _S( t)U( t )$ and $\tilde H_{SB} ( t ) \equiv U^\dagger ( t )H_{SB} U( t)$. If we write $x = \sqrt {\hbar /2m_t \Delta _t } ( {a_t  + a_t^\dagger  })$, and from Eq. (\ref{Inter_pict}), we obtain interaction terms similar to that of an ordinary (i.e. with mass and spring constant fixed) harmonic oscillator, except the replacement: (a) $\Delta _0 t \to \int_0^t {dt'} \Delta \left( t' \right)$ and (b) $m_0 \Delta _0  \to m_t \Delta _t$. Consequently, the resonating modes $\omega _k  \approx \Delta \left( t \right)$ would be time-dependent, and hence the friction ``coefficient" $\eta \left( t \right) \equiv \pi J\left( {\Delta _t } \right)/\Delta _t$, where $J\left( \omega  \right) \equiv \sum\nolimits_k {g_k^2 \delta \left( {\omega _k  - \omega } \right)}$, would also be a function of time i.e., with a classical equation of motion (neglect the frequency renormalization):
\begin{equation}\label{damped}
m\left( t \right)\frac{{d^2 }}{{dt^2 }}\left\langle x \right\rangle  + \eta \left( t \right)\frac{d}{{dt}}\left\langle x \right\rangle  + k\left( t \right)\left\langle x \right\rangle  = 0 \quad .
\end{equation}
The exception is the ohmic case, where $J\left( \omega  \right) \propto \omega$ and hence $\eta \left( t \right) = \eta _0$ is independent of the variation in the mass and the spring constant (e.g. RLC circuit). Finally, the equation for $n\left( t \right) = Tr\{ {a_0^\dagger  a_0 \tilde \rho _S \left( t \right)}\}$ is obtained from the master equation:
\begin{equation}\label{nt}
\frac{d}{{dt}}n\left( t \right) =  - \gamma \left( t \right)\left( {n\left( t \right) - N\left( t \right)} \right) \quad,
\end{equation}
where $N\left( t \right) \equiv 1/\left( {e^{ \Delta ( t ) / k_B T}  - 1} \right)$. This is the key result of this paper, since the performance of AQC is determined entirely by $n(t)$. Note that even for the case of ohmic damping, the relaxation rate $\gamma \left( t \right) \equiv \eta \left( t \right)/m\left( t \right)$ is in general time dependent. With the initial condition $n\left( 0 \right) = 0$, this equation can be solved numerically to obtain the ground state occupation $P_g$ at time $t$. 

Although in our model the time dependence for the energy gap is completely arbitrary, for the purpose of understanding the structure of the thermal excitation we assume that the gap has a Landau-Zener type variation:
\begin{equation}\label{Delta}
\Delta \left( t \right) = \sqrt {\Delta _{\max }^2 \left( {1 - t/\tau _* } \right)^2  + \Delta _{\min }^2 } \quad,
\end{equation}
where for $\Delta _{\max } \gg \Delta _{\min }$, $\Delta \left( 0 \right) = \Delta \left( {2\tau _* } \right) \approx \Delta _{\max }$, and $\Delta \left( {\tau _* } \right) = \Delta _{\min }$. Except near the region $t \approx \tau _0$, the rate of change of the energy gap is $V_S  \equiv \Delta _{\max } /\tau _*$. For simplicity, we shall consider the ohmic case only and assume that $m\left( t \right) = m_0$ is time-independent, which makes $\gamma$ time-independent as well. 

The following examples are chosen to demonstrate respectively that: (I) when thermalization is important (i.e. $\Delta \left( t \right) \le k_B T$), the computation speed $V_S$ needs to be fast, compared with the ``natural" speed of the bath $V_B  \equiv \gamma k_B T$. We quantify this by defining $R \equiv V_B /V_S  = \gamma k_B T\tau _* /\Delta _{\max }$. (II) After passing through the gap minimum, when the energy gap is larger than the temperature, i.e. $\Delta \left( {t > \tau _* } \right) > k_B T$, relaxation towards the ground state (increasing $P_g$) has the simple $e^{ - \gamma t}$ dependence, and in contrast with that in Ref.\cite{Amin08}, does not depend on $\Delta _{\min }^2 /\Delta _{\max }$ in the exponent. (III) This toy model qualitatively agrees with numerical calculation based on more realistic Hamiltonian.    

\begin{figure}[t]
\centering
\includegraphics[width=8.5cm]{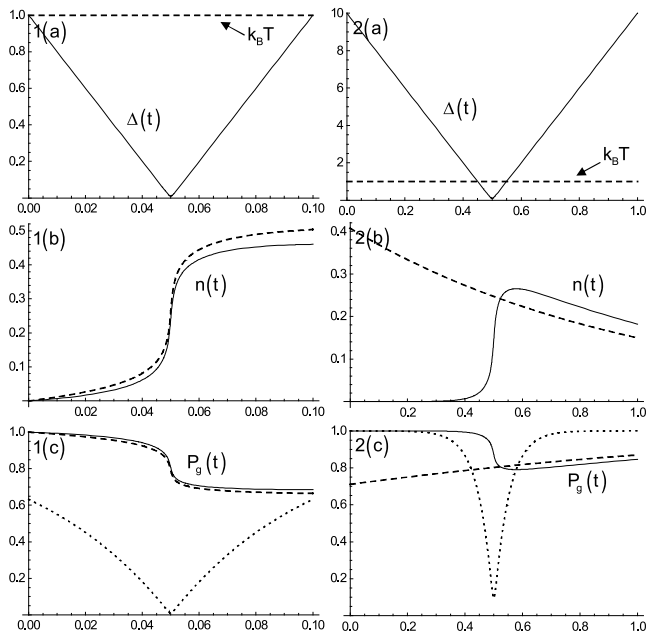}
\caption{Simulation of AQC with harmonic oscillator under thermal noise. The $x$-axis is rescaled time $ \gamma t$. 1(a) and 2(a), in unit of $k_B T$, show the energy profiles Eq. (\ref{Delta}) for two classes of AQC (but with the same $R$). 1(b) and 2(b) show the dynamics of the mean excitation of the oscillator. The solid lines are numerical integration from Eq. (\ref{nt}), and the dashed lines are analytic approximations Eq. (\ref{nt1}) and Eq. (\ref{nt2}). 1(c) and 2(c) are the corresponding ground state probabilities obtained from 1(b) and 2(b), with the relation Eq. (\ref{compact_Pg}). For comparison, the dotted lines are the thermal equilibrium values of $P_g  = 1 - e^{ - \beta \Delta}$.}
\label{fig:}
\end{figure}

\emph{Example I ---} Consider the case where ${\Delta _{\max }  \le k_B T}$ and $\Delta _{\min }  \ll k_B T$. It is possible to approximate $N\left( t \right) \approx k_B T/\Delta \left( t \right)$. Substitute this into Eq. (\ref{nt}), and with $\gamma \left( t \right) = \gamma$, we have 
\begin{equation}\label{nt1}
n\left( t \right) = \frac{{\gamma k_B T}}{{\Delta _{\max } }}e^{ - \gamma t} \int_0^t {ds\frac{{e^{\gamma s} }}{{\sqrt {\left( {1 - s/\tau _* } \right)^2  + \varepsilon ^2 } }}} \quad, 
\end{equation} 
where $\varepsilon  \equiv \Delta _{\min } /\Delta _{\max } \ll 1$. For $\gamma \tau _*  < 1$, the integrand is dominated near $s \approx \tau _*$. Taking $e^{\gamma s}  \to e^{\gamma \tau _* }$ and integrating explicitly, we have 
\begin{equation}
n\left( t \right) = \lambda(t) Re^{ - \gamma \left( {t - \tau_* } \right)} \quad, 
\end{equation}
where $\lambda(t)  \equiv \ln 2 - \ln [ {\sqrt {\varepsilon ^2  + \left( {1 - t/\tau_* } \right)^2 }  + \left( {1 - t/\tau_* } \right)}]$, and $R = \gamma k_B T\tau _* /\Delta _{\max }$ as defined above. Figure \ref{fig:} shows that this expression for $n(t)$ is in good agreement with the result by direct numerical integration for $n(t)$. From Eq. (\ref{compact_Pg}), we conclude that the thermal effect is not important even if $\Delta \left( t \right) \le k_B T$ provided that $\gamma \tau _*  \ll 1$. More precisely, we require $R \ll 1$, or $V_S \gg V_B$. This minimal speed limit for AQC could not be seen by the two-level approximation \cite{Amin08}.

\emph{Example II ---} Here we consider the possibility of relaxation after passing through the minimum. In other words, we consider $n\left( t \right)$ when $\Delta \left( {t > \tau _* } \right) > k_B T$, while $\Delta _{\min }  < k_B T$.  This situation should not be very common for AQC, as it suggests that thermalization from the heat bath would yield better performance. To start, we could not invoke the same approximation as in example I. However, as long as $R < 1$, $N\left( t \right)$ is still sharply peaked at $t \approx \tau _*$. Base on this observation, skipping the details, we obtain an approximate solution which is valid \textit{only} for $t>\tau_*$
\begin{equation}\label{nt2}
n\left( {t > \tau _* } \right) = \kappa Re^{ - \gamma \left( {t - \tau _* } \right)} \quad,
\end{equation}
where $\kappa  \equiv 2a + 2\ln \left[ {k_B T/\Delta _{\min } \left( {1 - R} \right)} \right]$ and $a \equiv \int_0^\infty  e^{ - x} \ln \left( {2x} \right)dx =  0.116$. Figure \ref{fig:} shows that this expression qualitatively agrees with the direct numerical integration for $n(t)$ for $t > \tau _*$. Again, we conclude that $R$ plays an important role for determining the performance of AQC. It is suggested \cite{Amin08} that the combination $\Delta _{\min }^2 /\Delta _{\max }$ would be important for the relaxation process in the two-level approximation. We have tested with different ratios of $\Delta _{\min } /\Delta _{\max }$ (while keeping $\Delta _{\min }  \ll \Delta _{\max }$), but we did not find explicit dependence of $\Delta _{\min }$ in the exponent of $n(t)$.

\emph{Example III ---} So far we have compared our results with that from the two-level approximation. Would a realistic Hamiltonian (with non-uniform distribution of energy gaps) for AQC, in some sense, look like a harmonic oscillator (uniform gaps)? If yes, then based on the results above, it may be possible to approximate the final ground state occupation by the formula:
\begin{equation}\label{fitting}
P_g  \approx \frac{1}{{1 + \alpha R}} \quad,
\end{equation}
where $R \equiv \gamma k_B T \tau_*/\left( {\Delta _{\max }  - \Delta _{\min } } \right)$ is generalized to include cases where $\Delta _{\min } /\Delta _{\max }$ is not negligible, and $\alpha$ is a fitting parameter. For example I, at $t=2\tau_*$, $\alpha  = 2e^{ - \gamma \tau_* } \ln \left( {2\Delta _{\max } /\Delta _{\min } } \right)$, and for example II, $\alpha  = \kappa e^{ - \gamma \tau _* }$. The former does not depend on $T$, and the latter depends on $T$ weakly (logarithmically).

In Ref.\cite{Childs01}, an algorithm solving the so-called ``three-bit exact cover" (EC3) problem, in which the energy gap (here taken as $\Delta \left( t \right)$) between the ground and first excited state of $H_S \left( t \right)$ is significantly larger than the rest when $\Delta \left( t \right) = \Delta _{\min }$. We extract, from FIG 2 of that paper, the final probability $P_g$ and $R$, and estimate the corresponding $\alpha$ by the relation in Eq. (\ref{fitting}), as suggested by our harmonic oscillator model. The results are shown in Table {\ref{AQC}. We found that both sets of data are consistent with the conjecture that $P_g$ decreases with $R$. For data I, the fluctuation for the value of $\alpha$ is relatively small (about 20\% from the mean), and the data point for high temperature case ($k_B T/\Delta _{\max }=10$) deviates significantly with the rest. This is anticipated from our experience in examples I and II. For data II, the fluctuation is relatively larger (about 40\% from the mean). This may due to that the ratio $\Delta _{\min } /\Delta _{\max }  = 0.425$ is a bit too large for our simple formula Eq. (\ref{fitting}) to be accurate. We conclude that this harmonic oscillator model may provide reasonable estimation for some realistic AQC problems.

For AQC involving many more degrees of freedom, it may (either computationally or experimentally) be challenging to obtain the eigenenergy spectrum. However, this model can still be applicable. We first determine the $\alpha$ and $R$ for an AQC with a relatively small problem size $n$ (i.e. what was done in Table \ref{AQC}). We then gradually increase $n$ and determine the (average) scaling $\Gamma _n < 1$ of the energy spectrum (typically the first few lowest energy states are enough). Then, to estimate the same AQC (under the same temperature) with large $N$, the harmonic oscillator model suggests the ground state probability be given by the formula Eq. (\ref{fitting}) with the replacement $\Delta _{\max }  \to \Gamma _N \Delta _{\max }$ (assuming $\tau_*$ and $\gamma$ are fixed).
 
\begin{table}[t]
\caption{\label{AQC} Simulation of AQC with Harmonic Oscillator. The data  ($P_g$ and $R$), taken from the numerical simulation (FIG 2) of Childs \textit{et al}. \cite{Childs01}, are fitted ($\alpha$ being the fitting parameter) with the formula in Eq. (\ref{fitting}) suggested by the harmonic oscillator model. The standard deviation (excluding the last data point $k_B T = 10$) of $\alpha$ for data I (II) is about $20\%$ ($40\%$) from the mean value 1.68 (0.81).}
\begin{ruledtabular}
\begin{tabular}{c|c c c c c c}
 & $k_B T$ \footnotetext{$k_B T$ and $\Delta _{\min }$ are in unit of ${\Delta _{\max } }$.} & $1/10$ & $1/2$ & $1$ & $2$ & $10$\\
\hline
Data I: & $P_g$ & 0.79 & 0.53 & 0.30 & 0.15 & 0.08\\
$\Delta _{\min }  {=} 0.301$ & $R$ & 0.14 & 0.72 & 1.43 & 2.86 & 14.3\\
 & $\alpha$ & 1.86 & 1.24 & 1.63 & 1.98 & 0.80\\
\hline
Data II: &  $P_g$ & 0.89 & 0.70 & 0.42 & 0.19 & 0.08\\
$\Delta _{\min }  {=} 0.425$ & $R$ & 0.17 & 0.87 & 1.74 & 3.48 & 17.4 \\
 & $\alpha$ & 0.71 & 0.49 & 0.79 & 1.23 & 0.66\\
\end{tabular}
\end{ruledtabular}
\end{table}

\emph{Conclusions ---} We have introduced a harmonic oscillator model for a quantitative study on the effect of thermal noise on AQC. For ohmic damping, we showed that AQC is considered \textit{fast}, if the combination $R \equiv \gamma k_B T\tau _* /\Delta _{\max }$ is small. This model suggests a simple relation for estimating the fidelity (cf. Eq.(\ref{fitting})) of general AQC; this relation qualitatively agrees with the previous numerical simulation. This model can also be verified with quantum RLC circuits, and therefore can act as a test bed for future theoretical and experimental investigation on AQC.  



\begin{acknowledgments}
M.H.Y acknowledges the support of the NSF grant EIA-01-21568 and the Croucher Foundation, and thanks  R. Laflamme for the hospitality of the Institute for Quantum Computing where part of this work is done. M.H.Y also thanks Jonathan Baugh, Andrew Childs and Tzu-Chieh Wei for valuable discussions, and especially A. J. Leggett for comments and criticisms.
\end{acknowledgments}



\end{document}